\documentclass[letter]{aa} 
\usepackage{graphicx}
\usepackage{txfonts}
\begin{document}
   \title{ULAS~J141623.94$+$134836.3 - a faint common proper motion 
          companion of a nearby L dwarf}

\subtitle{Serendipitous discovery of a cool brown dwarf in UKIDSS DR6}

   \author{R.-D. Scholz
          }
   \institute{Astrophysikalisches Institut Potsdam,
              An der Sternwarte 16, 14482 Potsdam, Germany\\
              \email{rdscholz@aip.de}
             }

   \date{Received 15 January, 2010; accepted 27 January, 2010}


  \abstract
   {}
   {New near-infrared large-area sky surveys (e.g. UKIDSS, CFBDS, WISE) 
    go deeper than 2MASS and aim at
    detecting brown dwarfs lurking in the Solar neighbourhood 
    which are even fainter than the latest known T-type
    objects, so-called Y dwarfs.
   }
   {Using UKIDSS data, we have found a faint brown dwarf candidate 
    with very red optical-to-near-infrared but extremely 
    blue near-infrared colours next to the recently discovered nearby
    L dwarf SDSS~J141624.08$+$134826.7. We check if the two objects 
    are co-moving by studying their parallactic and proper motion
    and compare the new object with known T dwarfs.
   }
   {The astrometric measurements are consistent
    with a physical pair ($sep$$\approx$75~AU)
    at a distance $d$$\approx$8~pc. The extreme colour 
    ($J$$-$$K$$\approx$$-$1.7) and absolute magnitude 
    ($M_J$=17.78$\pm$0.46 and $M_K$=19.45$\pm$0.52)
    make the new object appear
    as one of the coolest (T$_{eff}$$\approx$600~K)
    and nearest brown dwarfs, 
    probably of late-T spectral type and possibly
    with a high surface gravity (log $g$$\approx$5.0). 
   }
   {}


   \maketitle
%

\section{Introduction}

One of the open questions of low-mass star formation is
the ratio of successful/failed star formation processes.
In other words, is the Solar neighbourhood populated by
as many cool brown dwarfs as red dwarf stars? 
Several new near-infrared surveys like  
UKIDSS\footnote{The UKIDSS project is defined in Lawrence et al.~(\cite{lawrence07}).
UKIDSS uses the UKIRT Wide Field Camera (WFCAM; Casali et al.~\cite{casali07})
and a photometric system described in Hewett et al.~(\cite{hewett06})
which is in the Mauna Kea Observatories (MKO) system (Tokunaga et 
al.~\cite{tokunaga02}).
The pipeline processing and science archive are described
in Irwin et al.~(\cite{irwin10}) and Hambly et al.~(\cite{hambly08}).},
CFBDS (Delorme et al.~\cite{delorme08b}) and 
WISE (Wright~\cite{wright08}; Mainzer et al.~\cite{mainzer09})
try to answer this question by going deeper than
the Two Micron All Sky Survey (2MASS; Skrutskie et al.~\cite{skrutskie06})
to detect a new class of ultracool brown dwarfs, so-called Y dwarfs.

The appearance of ammonia absorption in the near-infrared
spectra is beeing discussed as a criterion for the new Y spectral type
(Burningham et al.~\cite{burningham08}; Delorme et al.~\cite{delorme08b}).
Whereas the latest-type (coolest) objects discovered in the
2MASS are of spectral type T8 (Burgasser et al.~\cite{burgasser02}; 
Tinney et al.~\cite{tinney05}; Looper, Kirkpatrick, \& 
Burgasser~\cite{looper07}), a handful of even cooler 
(T$_{eff}$$\approx$500-600~K)
brown dwarfs (T8.5-T9) have already been discovered in UKIDSS
(Warren et al.~(\cite{warren07}; 
Burningham et al.~\cite{burningham08,burningham09}) and
CFBDS (Delorme et al.~\cite{delorme08a})
that do not look obviously different in their near-infrared 
spectra from late-type T  dwarfs. A unique 
Y dwarf has not yet been found and classified. 

In this letter, we describe a new cool brown dwarf, which
is probably a late-T dwarf with unusual properties, detected as
a wide companion of a nearby blue L dwarf.

\section{Identification of a faint object with unusual colors 
near the blue L dwarf SDSS~J141624.08$+$134826.7}

\begin{figure}
   \centering
   \includegraphics[height=50mm,angle=0]{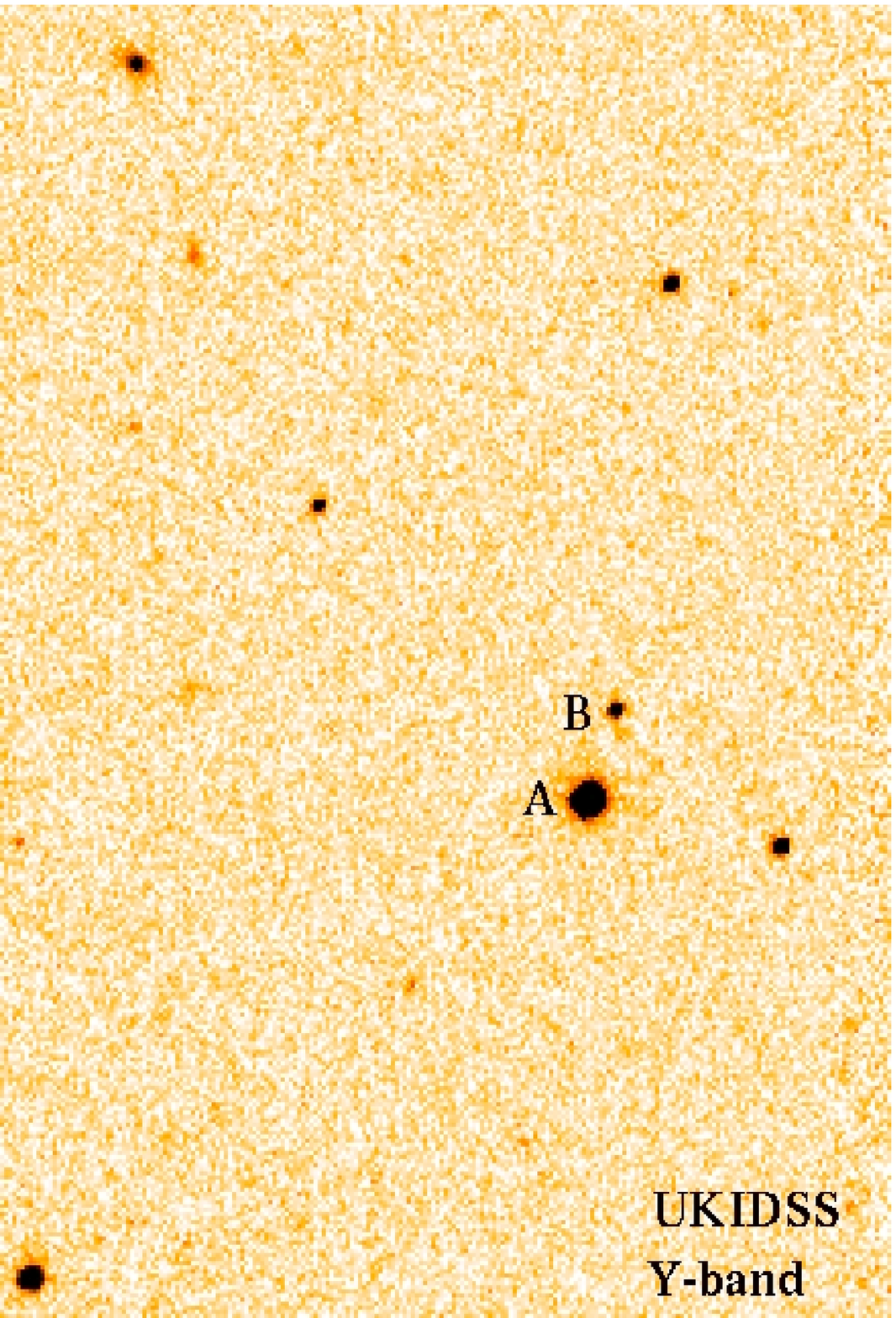}
   \includegraphics[height=50mm,angle=0]{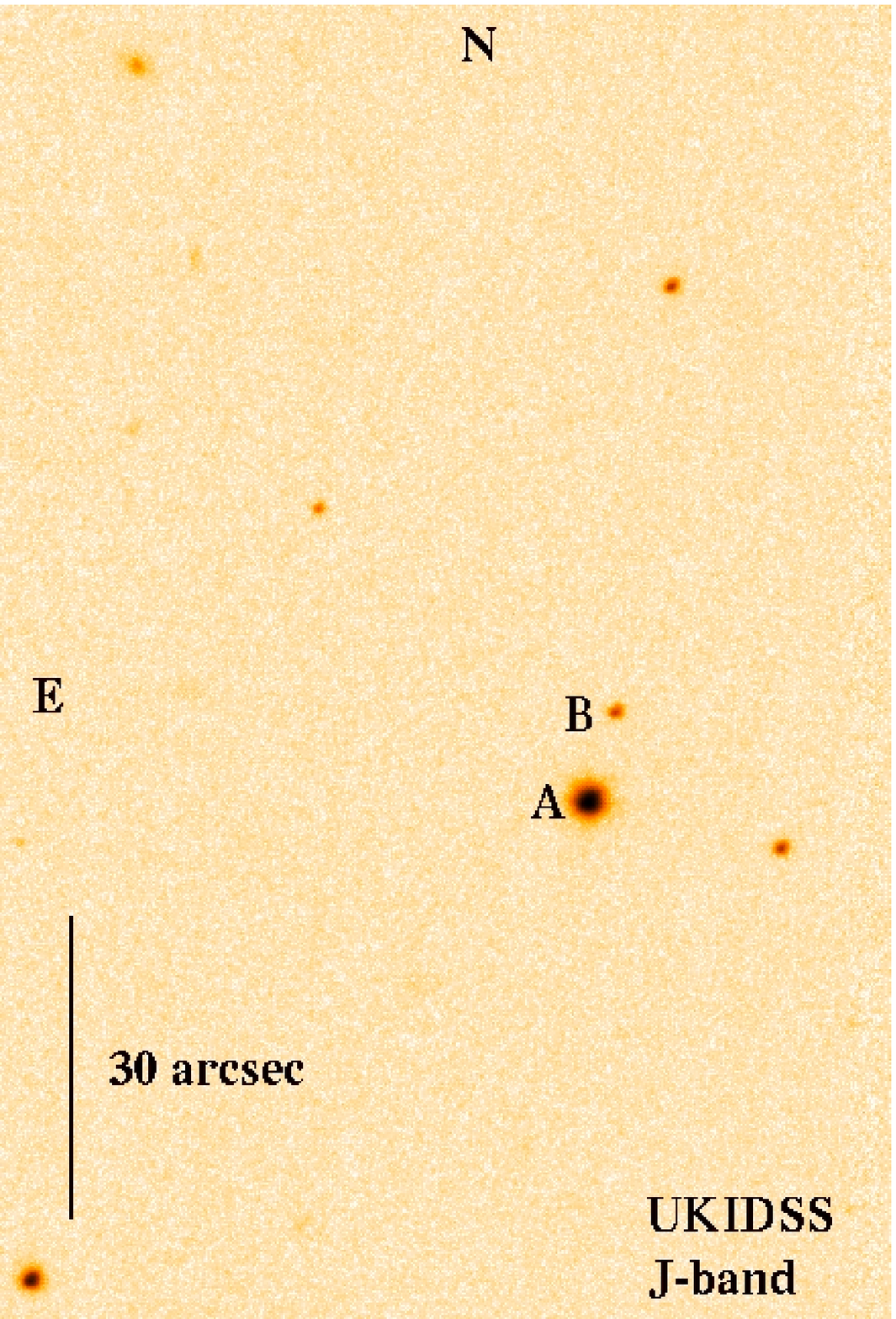}
   \includegraphics[height=50mm,angle=0]{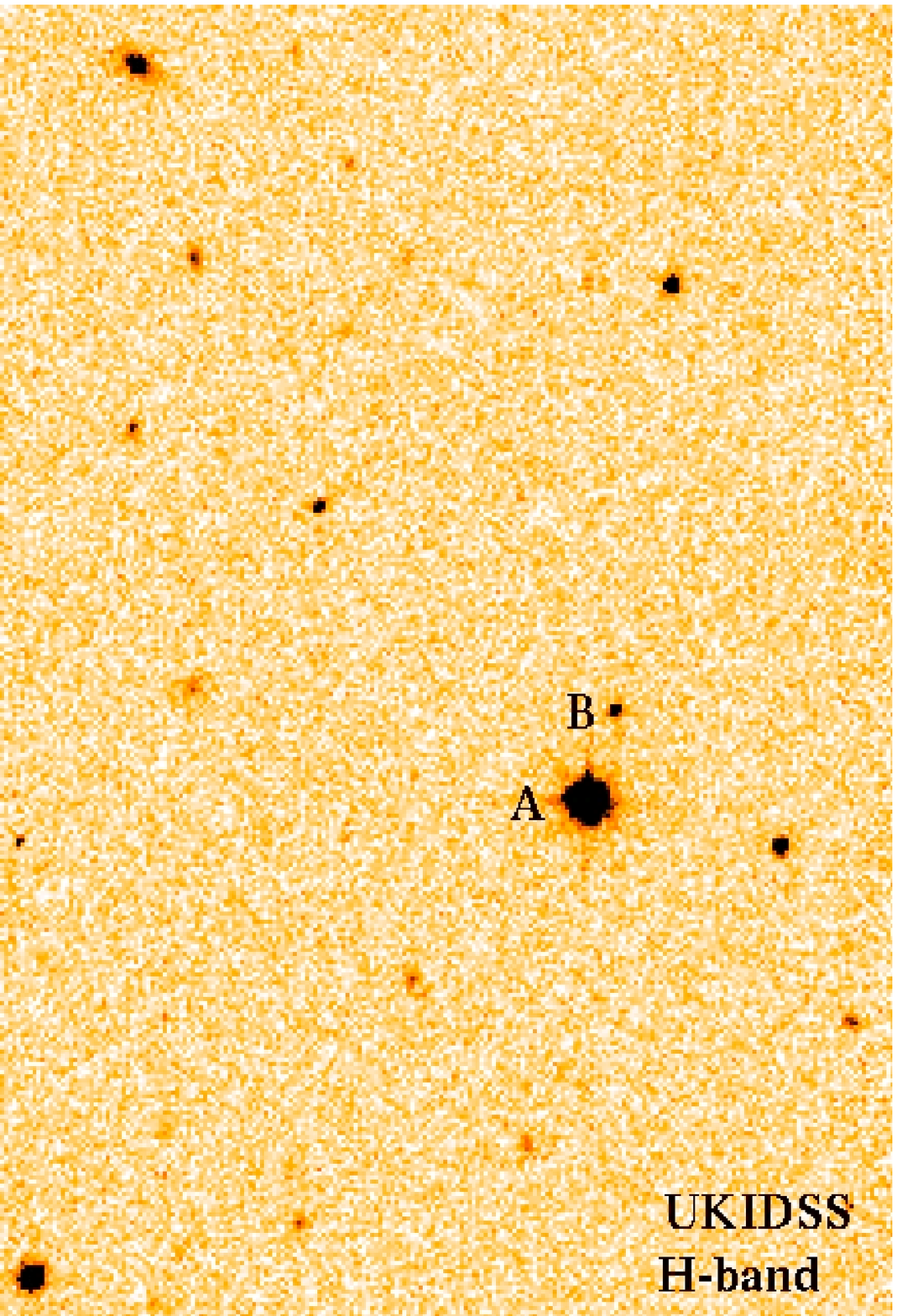}
   \includegraphics[height=50mm,angle=0]{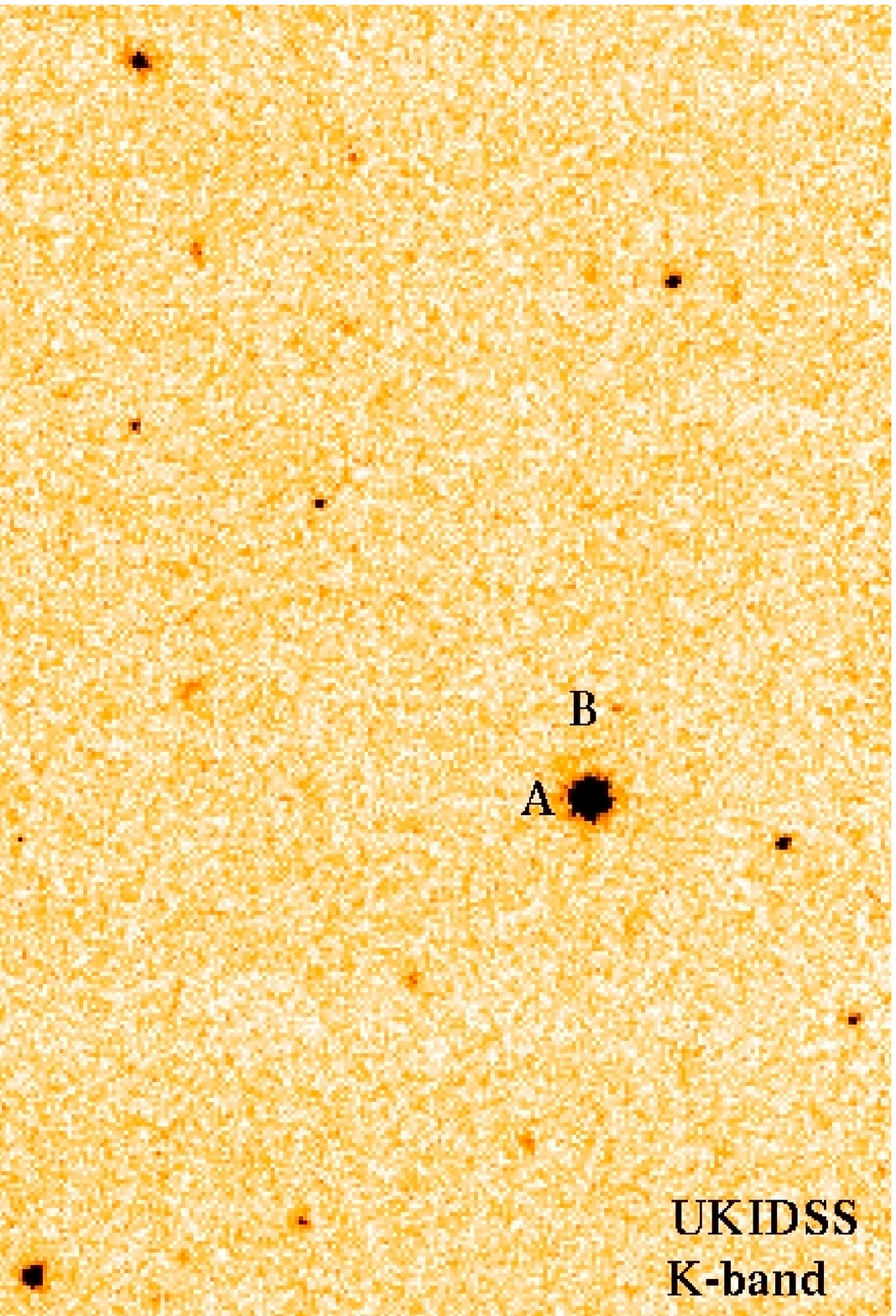}
      \caption{UKIDSS $YJHK$ images of objects A and B.
               Image scale
               and orientation (north is up, east to the left)
               are shown in the J-band image.
              }
         \label{fig_fcharts}
   \end{figure}

While inspecting the UKIDSS finding charts
around the recently discovered (Schmidt et al.~\cite{schmidt09}, herafter S09; 
Bowler, Liu \& Dupuy~\cite{bowler09}, hereafter B09) nearby blue L6 dwarf 
SDSS~J141624.08$+$134826.7 (hereafter called object A), we found a fainter 
nearest neigbouring object with extreme colours,
separated by about 9.4~arcsec.
This object, ULAS~J141623.94$+$134836.3 (herafter called object B) 
was not detected in the SDSS DR7 (Abazajian et al.~\cite{abazajian09}), but
is well measured in the UKIDSS, where it has $Y$$-$$J$$\approx$$+$0.9
but a very blue near-infrared
colour of $J$$-$$H$$\approx$$-$0.3 and $J$$-$$K$$\approx$$-$1.7 
(Fig.~\ref{fig_fcharts}; Tab.~\ref{tab_photom}). 
Abazajian et al.~(\cite{abazajian09}) describe the completeness
limit of SDSS DR7 with a 95\% detection repeatability for point sources
at $u$=22.0, $g$=22.2, $r$=22.2, $i$=21.3, and $z$=20.5.
The non-detection of object B in SDSS DR7 hints at a very
red optical-to-near-infrared colour ($z$$-$$Y$$>$$+$2.3 and $z$$-$$J$$>$$+$3.1).
Using two available overlapping $z$-band FITS images (SDSS runs 3971 and 3996)
downloaded from the SDSS DR7, 
we were able to detect object B (for the astrometry see Tab.~\ref{tab_astm})
and measure its magnitude as $z_{3971}$=21.24$\pm$0.50 and
$z_{3996}$=21.02$\pm$0.39. The resulting mean colour indices are
$z$$-$$Y$=$+$2.97$\pm$0.32 and $z$$-$$J$=$+$3.87$\pm$0.32.

\begin{figure}
   \centering
   \includegraphics[width=60mm,angle=270]{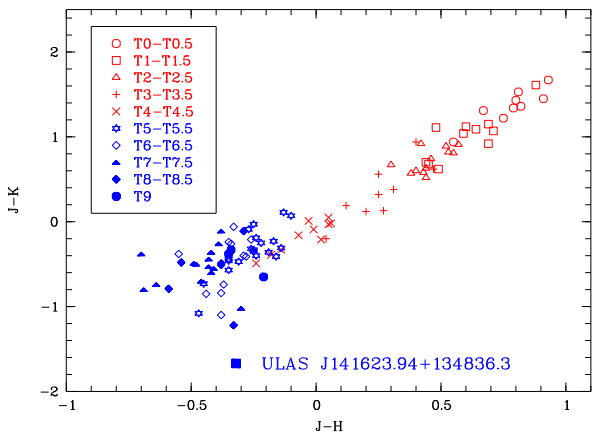}
      \caption{Near-infrared two-colour diagram $J$$-$$K$ vs. $J$$-$$H$
               (in MKO system) for all T dwarfs
               in Leggett et al.~(\cite{leggett10}) and the 
               new cool brown dwarf
               candidate ULAS~J141623.94$+$134836.3 (object B).
              }
         \label{fig_dwarch1}
   \end{figure}

Comparing the near-infrared colour indices of object B with those of
the known T dwarfs (Fig.~\ref{fig_dwarch1}), one can see that similar
moderately negative $J$$-$$H$ have been measured for the latest-type (T9)
but also for other mid- and late-type T dwarfs, whereas the extremely
large negative $J$$-$$K$ of object B clearly stands out against the
rest of the T dwarfs. Both colours are in the range typical of model
T and Y dwarfs but rule out a high-redshift quasar
(Hewett et al.~\cite{hewett06}). However, before further analysis
we need to check the physical association of object B with
object A and to confirm its distance.

%
\begin{table*}
\caption{SDSS DR7 $z$ (on the AB system) and UKIDSS DR6 $YJHK$ 
(on the Vega system using the MKO photometric system) photometry}
\label{tab_photom}
\centering
\begin{tabular}{l c c c c c c c  }
\hline\hline
object          & mean $z$       &  $Y$    & $J$     & $H$     & $K$     \\
                &(3971+3996)    &(2325240)&(2325252)&(1906043)&(1905923)\\ 
\hline
SDSS~J141624.08$+$134826.7 (object A) & 15.897$\pm$0.005 & 14.255$\pm$0.003 & 12.995$\pm$0.001 & 12.469$\pm$0.001 & 12.053$\pm$0.001 \\
ULAS~J141623.94$+$134836.3 (object B) & 21.13$\pm$0.32$^*$ & 18.162$\pm$0.027 & 17.259$\pm$0.017 & 17.581$\pm$0.030 & 18.933$\pm$0.244 \\
\hline
\end{tabular}

\smallskip

\scriptsize{
Notes: 
$^*$ - not detected in SDSS DR7 (see text).
$z$ magnitudes 
are mean values from two SDSS runs (given in brackets).
$YJHK$ magnitudes 
are {\it{aperMag3}} derived from the multi\-frames (given in brackets)
for point sources
(Dye et al.~\cite{dye06}). A second set of $YJHK$ 
measurements was not used due to the location of objects A and B
close to the edge of the frames.
}
\end{table*}

\begin{table}
\caption{Multi-epoch positions $\alpha, \delta$ (J2000.0): 14$^h$16$^m$...$^s$, $+$13\degr48\arcmin...\arcsec}   
\label{tab_astm}      
\centering            
\begin{tabular}{l l l l l l}          
\hline\hline  
\multicolumn{2}{c }{A} & \multicolumn{2}{c }{B} & epoch    & source    \\
\multicolumn{1}{c}{$^s$} & \multicolumn{1}{c}{\arcsec} & \multicolumn{1}{c}{$^s$} & \multicolumn{1}{c}{\arcsec} & yr &   \\
\hline
23.804   &19.88   &         &       &1954.338 &SSS $E$ \\
24.050   &25.48   &         &       &1994.338 &SSS $R$ \\
24.068   &25.85   &         &       &1997.521 &SSS $I$ \\
24.0847  &26.345  &         &       &2000.164 &2MASS \\
24.0886  &26.741  &23.886$^*$   &35.88$^*$  &2003.409 &SDSS 3971 \\
24.0859  &26.683  &23.916$^*$   &35.49$^*$  &2003.472 &SDSS 3996 \\
24.1305  &27.339  &23.9480  &36.268 &2008.189 &UKIDSS $H$ \\
24.1325  &27.355  &23.9466  &36.273 &2008.189 &UKIDSS $K$ \\
24.1260  &27.410  &23.9421  &36.319 &2008.362 &UKIDSS $Y$ \\
24.1254  &27.416  &23.9435  &36.334 &2008.362 &UKIDSS $J$ \\
\hline
\end{tabular}

\smallskip

\scriptsize{
Note:
$^*$ - not detected in SDSS DR7 (see text).
}
\end{table}

\section{Confirmation of common proper motion}

\begin{figure}
   \centering
   \includegraphics[width=55mm,angle=270]{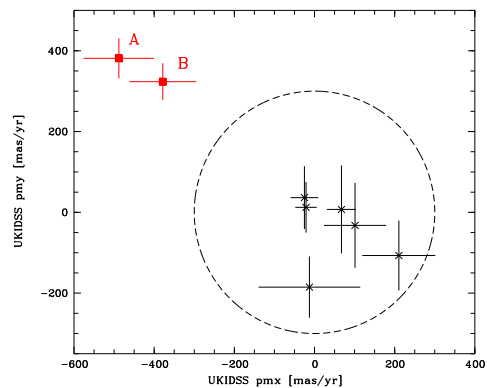}
      \caption{Short-term proper motions (from 2$\times$2 UKIDSS epochs 
               separated by 0.17 years) of objects A and B (filled squares,
               solutions 1-A and 1-B in Tab.~\ref{tab_pm})
               in comparison to those of field stars (crosses). 
               Typical proper motion errors of
               $\sigma$$\approx$75~mas/yr were achieved in both directions. 
               The dashed circle 
               represents a 4$\sigma$ significance level. 
              }
         \label{fig_ukidsspm}
   \end{figure}

For a first check of a possible common motion of objects A and B,
one can use the accurate UKIDSS data alone. There are two different
epochs for the $HK$ and $YJ$ observations, respectively 
(Tab.~\ref{tab_astm}). The corresponding multiframe numbers are
listed in Tab~\ref{tab_photom}. Short-term proper motions have been
determined from simple linear fitting over the four epoch positions
of objects A and B as well as of 6 field stars in their vicinity,
well-measured on the same multiframes (Fig.~\ref{fig_ukidsspm}).
Significant results, which agree within their errors, were obtained
for A and B (solution 1-A and 1-B in Tab.~\ref{tab_pm}). The common
short-term proper motion of A and B is a first strong hint on a
physical pair, but it is much larger and in a different direction
than the long-term proper motions of object A obtained by  
S09 and B09.
We will show that this discrepancy can be explained by the expected
common parallactic motion.

\begin{figure}
   \centering
   \includegraphics[width=60mm,angle=0]{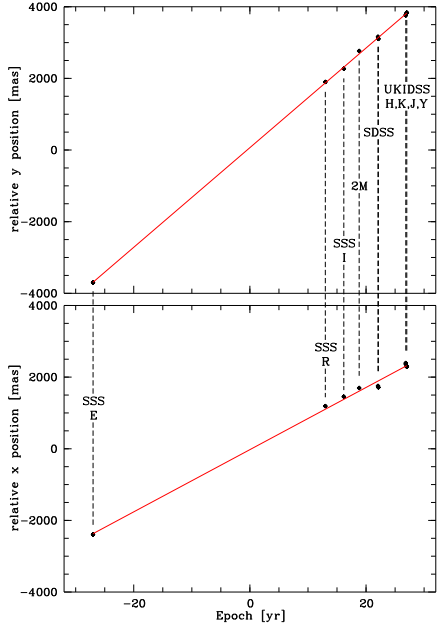}
      \caption{Linear proper motion fit for object A 
               (solution 3-A).
              }
         \label{fig_linpmA}
   \end{figure}

Using two $z$-band SDSS images containing object A, we were
able to detect object B with the help of the ESO skycat tool
and the ''pick object'' option which is based on Gaussian fitting
(Tab.~\ref{tab_astm}). We think the reason why object B does not
appear in the SDSS DR7 is that it is $>$0.5~mag fainter than the already 
mentioned 95\% detection limit in $z$ and can not be detected in $ugri$,
where it should be much fainter than the corresponding limits.
Using now our two SDSS positions of object B together with its
four UKIDSS positions, we get again similar proper motions 
(solutions 2-A and 2-B in Tab.~\ref{tab_pm}), 
now also approaching the known long-term proper
motion of object A. The latter has been further improved
by us (Fig.~\ref{fig_linpmA}, solution 3-A in Tab.~\ref{tab_pm}) 
using all available epochs including
2MASS and SuperCOSMOS Sky Surveys (SSS; Hambly et al.~\cite{hambly01}) 
data (Tab.~\ref{tab_astm}). Note that S09
did not use UKIDSS, whereas B09 
missed the important old SSS $E$ epoch for their proper motion
solutions of object A.


The spectrophotometric distance of object A
is still very uncertain (8.0$\pm$1.6~pc according to S09
and 8.4$\pm$1.9~pc according to
B09), because the spectral type-absolute
magnitude relations are not yet well-determined for the class of
blue L dwarfs.
B09 mentioned a notable
parallactic motion of object A, but their trigonometric parallax
($\pi_{rel}=107\pm34$~mas) leads to a less accurate distance
($9.3\pm3.0$~pc) than the aforementioned spectrophotometric
distance estimates.

We have applied the software of Gudehus~(\cite{gudehus01})
for combined proper motion and parallax solutions.
In the full solution for
object A (solution 4-A), we made use of all 10 available epochs assigning 
the following uncertainties to the $\alpha,\delta$ given in
Tab.~\ref{tab_astm}: 70~mas for UKIDSS and SDSS, 100~mas for
2MASS, 150~mas for SSS $I$-band, 200~mas for SSS $R$-band, and
250~mas for the SSS measurement of the old $E$ plate
(expected colour-dependent systematic errors in the different
$\alpha,\delta$ are much smaller and have been neglected).
As an alternative, we used only the most accurate data (UKIDSS
and SDSS) and the proper motion obtained in solution 3-A as a
fixed input parameter in fitting only the parallactic motion
of object A (solution 5-A in Tab.~\ref{tab_pm}) and object B 
(solution 5-B). In the latter case we assigned uncertainties of 200~mas
to our SDSS $\alpha,\delta$ measurements.

Our preferred solution 
for object A
(4-A) gives a proper motion
nearly identical to the linear fit (3-A) and provides a parallax leading
to a distance of 7.9$\pm$1.7~pc in perfect/good agreement with the
spectrophotometric distances of S09/B09, respectively.
Its accuracy is also comparable with that of the spectrophotometric
estimates. However, the full range of the parallaxes$\pm$errors 
in solutions 4-A, 5-A, and 5-B implies a larger uncertainty for the system.
Figure~\ref{fig_ABplx} shows that the short-term (UKIDSS only) proper motion 
of both objects (solutions 1-A and 1-B) is well-explained by their common
parallactic motion (the parallax results of solutions 5-A and 5-B agree
within their errors).

\begin{table}
\caption{Proper motion and parallax solutions for objects A and B}
\label{tab_pm}   
\centering        
\begin{tabular}{c r r r}          
\hline\hline
Solution  & $\mu_{\alpha}\cos{\delta}$ & $\mu_{\delta}$ & $\pi_{rel}$ \\
     & mas/yr & mas/yr & mas \\
\hline
1-A  &  $-$488.1$\pm$87.8  & $+$381.3$\pm$49.4  & \\
1-B  &  $-$378.6$\pm$83.3 & $+$323.5$\pm$45.7   & \\
2-A  &  $+$123.9$\pm$10.1  & $+$132.7$\pm$28.3   & \\
2-B  &  $+$138.3$\pm$05.8   & $+$126.7$\pm$25.4   & \\
3-A  &   $+$86.8$\pm$02.1   & $+$139.1$\pm$00.9    & \\
{\bf 4-A}  & {\bf  $+$86.2$\pm$02.6}  &{\bf $+$138.8$\pm$02.6}   & {\bf 127.0$\pm$27.0} \\
5-A  &                     &                     &  153.7$\pm$20.5  \\
5-B  &                     &                     &  104.4$\pm$45.5  \\
\hline

\hline
\end{tabular}

\smallskip

\scriptsize{
Linear fit using UKIDSS (1),
UKIDSS$+$SDSS (2), 
all data for A (3).
Combined proper motion and parallax solution for A (4).
Parallax solution using UKIDSS$+$SDSS and the previously
determined linear proper motion of object A as input (5).
}
\end{table}

   \begin{figure}
   \centering
   \includegraphics[width=88mm,angle=0]{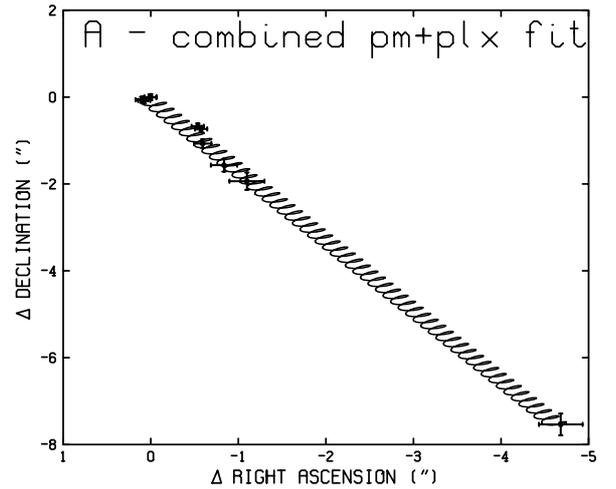}
      \caption{Combined proper motion and parallax solution with all
               available epochs for object A (solution 4-A). 
               Zero point is the UKIDSS $J$ epoch.
              }
         \label{fig_Apmplx}
   \end{figure}

   \begin{figure}
   \centering
   \includegraphics[width=44mm,angle=0]{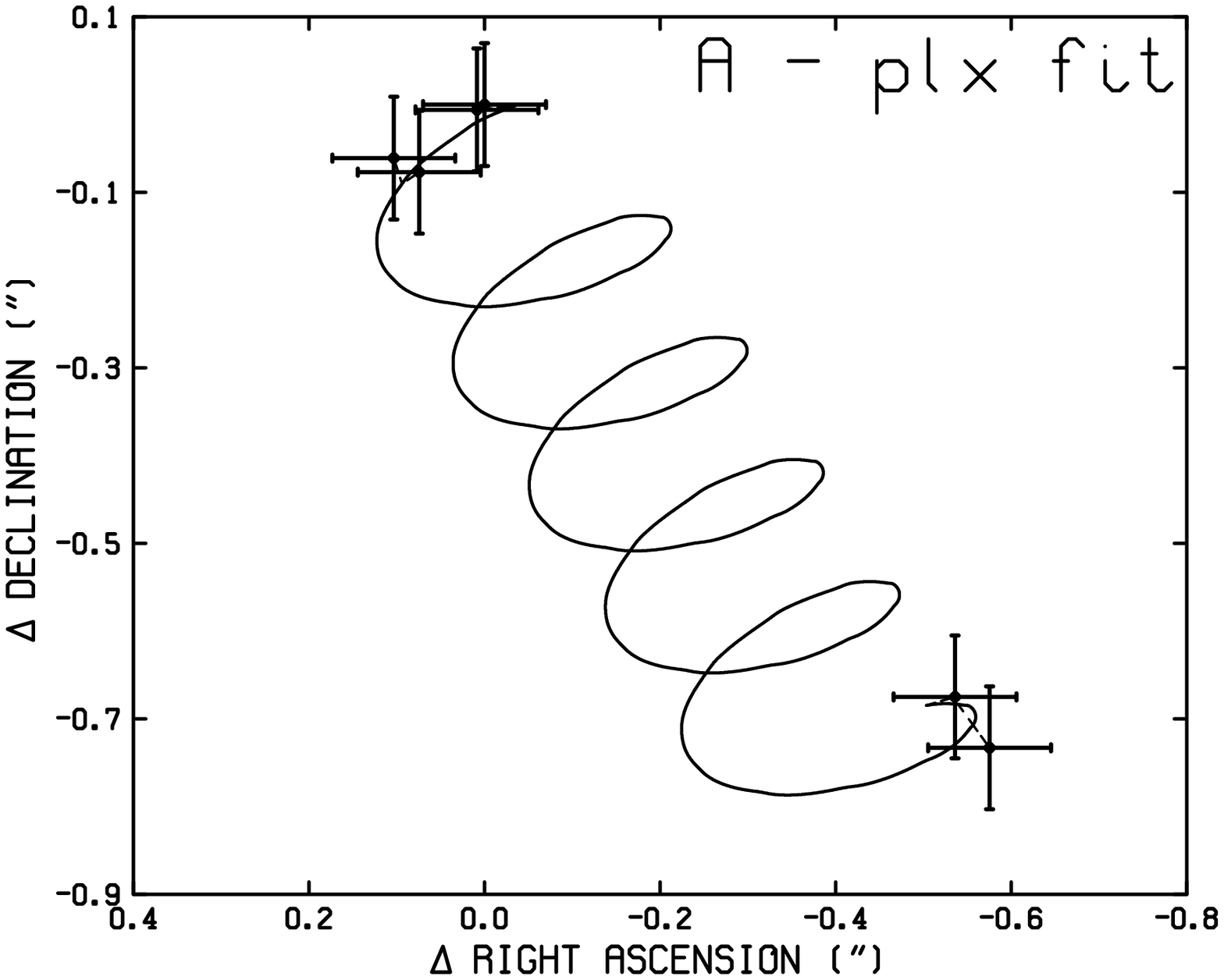}
   \includegraphics[width=44mm,angle=0]{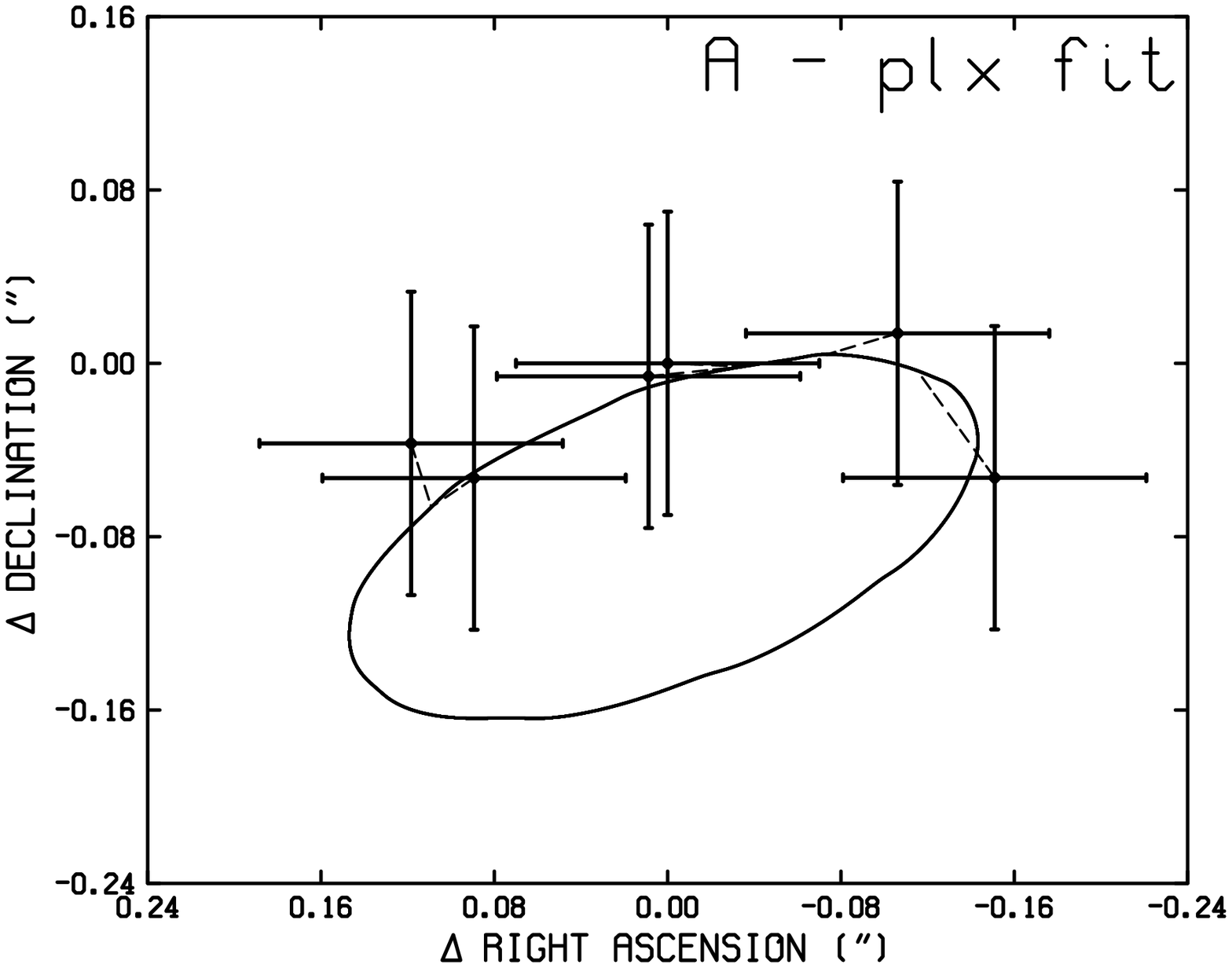}
   \includegraphics[width=88mm,angle=0]{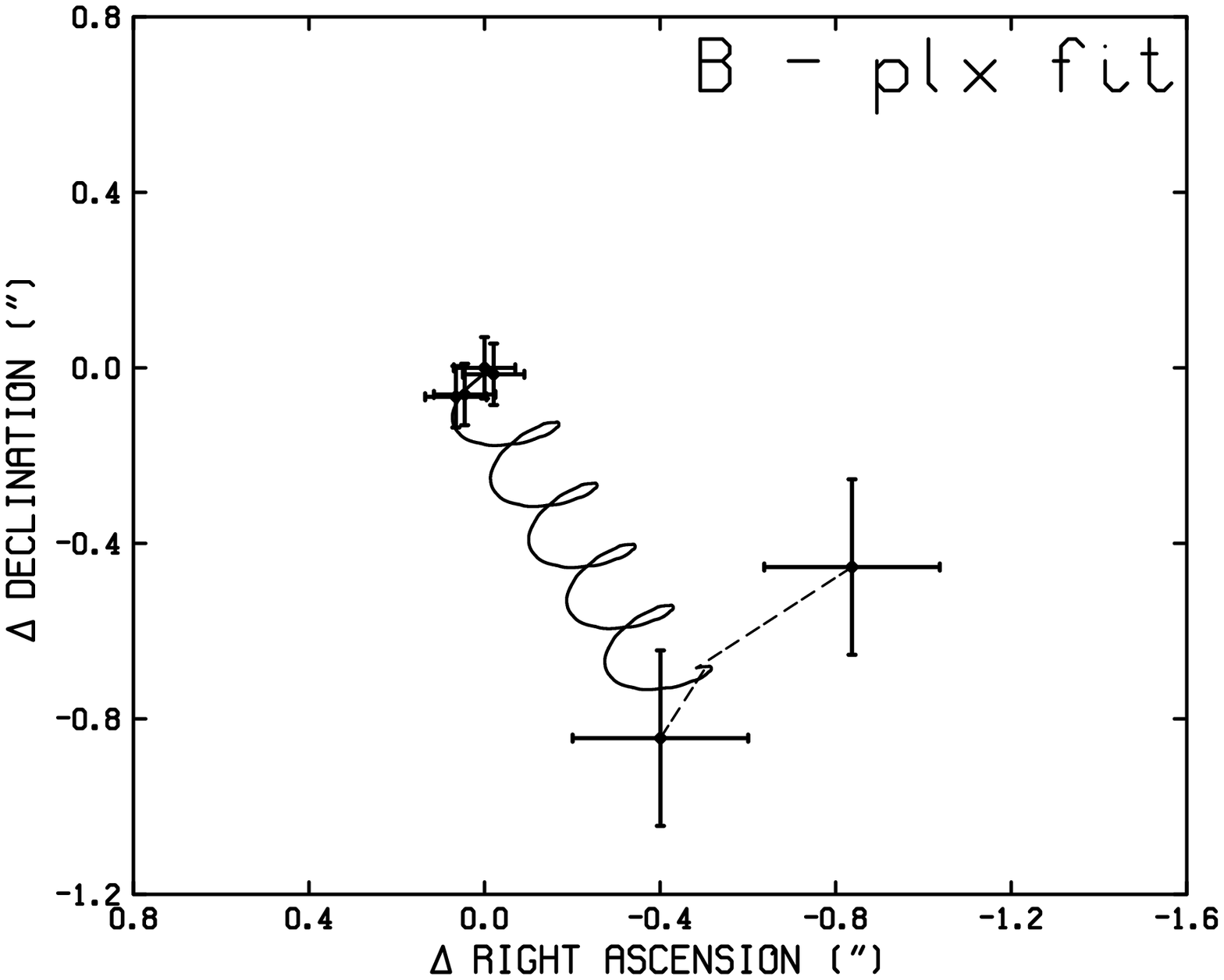}
      \caption{Parallax solution for objects A (top, solution 5-A;
               top, right shows the parallax fit with
               the proper motion removed) 
               and B (bottom, solution 5-B) using UKIDSS$+$SDSS 
               and the known long-term proper motion 
               (from solution 3-A) as fixed input parameter.
              }
         \label{fig_ABplx}
   \end{figure}

   \begin{figure}
   \centering
   \includegraphics[width=55mm,angle=270]{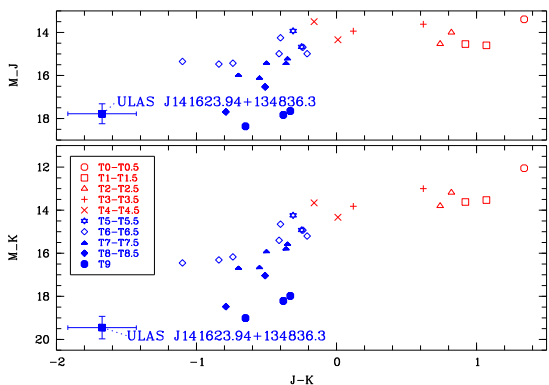}
      \caption{Absolute magnitudes $M_K$ (bottom) and $M_J$ (top)
               vs. $J$$-$$K$ colour 
               (in MKO system)
               for all T dwarfs with measured trigonometric parallaxes
               as listed in Leggett et al.~(\cite{leggett10})
               and 
               for ULAS~J141623.94$+$134836.3 (object B) with error
               bars obtained from using the parallax of object A (solution 4-A) 
               and the magnitude errors of object B (Tab.~\ref{tab_photom}). 
              }
         \label{fig_dwarch2}
   \end{figure}

\section{Conclusions and discussion}~\label{res}

We have discovered a faint common proper motion companion (object B) of a
blue nearby L6 dwarf (object A). Based on the astrometric measurements, 
which are
consistent with a wide binary (projected physical separation 75~AU)
at a distance of about 8~pc, and on the
accurate near-infrared colours placing object B 
at the end of 
the T dwarf sequence (Figs.~\ref{fig_dwarch1},\ref{fig_dwarch2}), 
we conclude that this object 
is one of the coolest known brown dwarfs,
probably with a late-T spectral type.

The latest-type brown dwarfs with trigonometric parallaxes available
are the T8.5 dwarfs Wolf 940B at a distance of 12.5$\pm$0.7~pc 
(= ULAS~J214638.83-001038.7; 
Burningham et al.~\cite{burningham09}; with a parallax measurement
for the primary Wolf 940A by Harrington \& Dahn \cite{harrington80}) and 
ULAS~J003402.77$-$005206.7 (Warren et al.~\cite{warren07};
Smart et al.~\cite{smart09}) at a distance of
12.6$\pm$0.6~pc. Gelino et al.~(\cite{gelino09})
list only one more T8.5 (ULAS~J123828.51$+$095351.3;
Burningham et al.~\cite{burningham08}) and two T9 dwarfs
(ULAS~J133553.45$+$113005.2; Burningham et al.~\cite{burningham08}
and CFBDS~J005910.90$-$011401.3; Delorme et al.\cite{delorme08a})
still lacking trigonometric parallaxes. However, their 
spectrophotometric estimates hint at distances of (slightly) more than 8~pc.
Object B is by 0.5-1.7 magnitudes brighter in the $J$- and $H$-band than
the above mentioned five objects. In particular, the possibly nearest of the
objects, the T9 dwarf ULAS~J133553.45$+$113005.2 at 8-12~pc according to 
Burningham et al.~(\cite{burningham08}), is 0.5-0.6~mag fainter than
object B in the $YJH$-bands whereas it is about 0.5~mag brighter than
object B in the $K$-band. Adopting the mean distance of 10~pc for
ULAS~J133553.45$+$113005.2 and 8~pc for object B, their absolute $YJH$
magnitudes are comparable, whereas object B is fainter in $M_K$
(Fig.~\ref{fig_dwarch2}).
Therefore, we think that object B 
is probably the nearest 
among the latest-type 
brown dwarfs offering excellent opportunities for follow-up
observations.

With an $H$$-$$K$$\approx$$-$1.35 and $M_H$$\approx$18.1, object B
falls outside Fig.~9 (top panel) in Leggett et al.~(\cite{leggett10}),
where these authors compare T dwarf observations with models. 
However, extrapolating
the model line with solar metallicity but high gravity (log $g$ = 5.0)
gives the best fit, possibly with a T$_{eff}$$\approx$600~K. Alternatively,
a slightly lower metallicity would also fit, but B09 excluded an
L subdwarf classification of object A, and the kinematics of the system
is clearly not typical of the Galactic halo or thick disk. The blue colour
of object A could also be caused by high 
surface gravity as discussed by Burgasser et al.~(\cite{burgasser08})
for the class of blue L dwarfs. If the high
gravity is correct, then the evolution models of Saumon \& Marley 
(\cite{saumon08}; their Fig.~4) show that the system is likely  
$\approx$5~Gyr old, and object B could have a mass of $\approx$30 Jupiters.
Further investigation
will show whether objects A$+$B represent a wide binary brown dwarf or
a much older analogue of the young low-mass star$+$massive planet
system 2MASS~1207$-$3932AB (Gizis~\cite{gizis02}; 
Chauvin et al.~\cite{chauvin04}).

\begin{acknowledgements}
Data from the UKIDSS 6th data release, the SDSS DR7, the 2MASS, and the 
SSS have been used for this work. We would like to thank Natasha Maddox 
for helpful advice on UKIDSS data and Hans Zinnecker for reading a first 
version of the manuscript.
We thank the referee, Dr Sandy Leggett, for
her prompt report which helped us to improve the paper.
\end{acknowledgements}

\end{document}